# Partially Ionized Plasma Physics and Technological Applications

Igor Kaganovich[1], and Michael Tendler [1,2*]
[1]Princeton Plasma Physics Laboratory, Princeton University, NJ 08543 USA; ikaganov@pppl.gov
[2]The Royal Institute of Technology, Stockholm (KTH), SE-100 44 Stockholm Sweden; mtendler@kth.se
[*]Correspondence: mtendler@kth.se.

**Abstract:** Partially ionized plasma physics has attracted a lot of attention recently due to numerous technological applications made possible by the increased sophistication of computer modelling, the depth of the theoretical analysis, and the technological applications to a vast field of the manufacturing for computer components. The partially ionized plasma is characterized by a significant presence of neutral particles in contrast to fully ionized plasma. The theoretical analysis is based upon solutions of the kinetic Boltzmann equation yielding the non-Maxwellian electron energy distribution function (EEDF) thereby emphasizing the difference with a fully ionized plasma. The impact of the effect on discharges in inert and molecular gases is described in detail yielding the complex nonlinear phenomena in plasma self-organization. A few examples of such phenomena are given including the non-monotonic EEDFs in the discharge afterglow in mixture of argon with the molecular gas $NF_3$; the explosive generation of cold electron populations in capacitive discharges, hysteresis of EEDF in inductively coupled plasmas. Recently, highly advanced computer codes were developed in order to address the outstanding problems of plasma technology. These developments are briefly described in general terms.

## 1. Introduction

Analysis of the low temperature plasma offers a lot of common features and similarities with high temperature plasmas relevant for the fusion research. In a partially ionized plasmas electrons are still hot enough to provide the energy required to dissociate molecules or ionize atoms and chemical radicals. While studying electrified gases in 1927, Irving Langmuir was influenced by the similarities between how blood plasma transports red and white blood cells and how an electrified gas carries around electrons and ions. This led him to adopt the term *plasma* for ionized gases. Thence, both quite different phenomena got the same name. When a gas is ionized, pairs of electrons and ions are created. The resulting plasma is like a "soup" of freely moving positive ions and negative electrons. The crucial point to bear in mind is that the total number of positively charged particles - ions is *approximately* equal to the total number of negatively charged particles - electrons and negative ions. The caveat (in italics) is crucial for understanding of both fully ionized and partially ionized plasma physics. The individual electrons and ions can be pulled closely or pushed apart by electrical and magnetic forces. However, electrons move faster due to their lower mass and higher mobility, and this movement can cause a separation of charge that is positive where the ions remain, and negative where electrons accumulate in peripheral plasma regions. Plasma remains *quasi-neutral*, implying that over large scales, the densities of positive and negative charges are nearly equal. This simple fact has profound ramifications for both partially and fully ionized plasmas. For example, it is the cause for the emergence of the self-consistent electric field in fully ionized plasmas that keeps negatively and positively charged particles together. The resulting electric field profile affects the quality of confinement exerted upon plasma by a magnetic field in fusion devices. Therefore, it is crucial for the impact on plasma turbulence. Indeed, it was demonstrated by numerous experiments that spatial gradients of the electric field shred multi-scale fluid eddies and result in the improved confinement regime in modern tokamaks [1,2,3]. In summary, both partially and fully ionized plasmas demonstrate many self-organizing features typical of a strongly non-linear system addressed within the framework of the chaos theory. These are the subject to the well-known non-linear phenomena in nature such as bifurcations of a steady state, the hysteresis of the timely evolution and the unfolding self-organization characterized by the power law behavior of the system [4,5,6,7].

The impact of an external magnetic field is another important factor affecting both fully and partially ionized plasmas, yet quite differently. Indeed, there is a big difference between fully ionized and partially ionized plasmas specified here. If plasma is fully ionized, increasing the power leads to heating and increasing electron or ion temperatures. (This approach is employed in fusion research in order to obtain relevant plasma parameters.) In contrast if plasma is partially ionized, additional power provides for an increase of the plasma density. The electron temperature is nearly fixed by the balance for rates of ionization and the loss of particles to surrounding walls. The averaged ionization frequency over a Maxwellian electron energy distribution function is the very non-linear function of the electron temperature. In partially ionized plasmas at high pressure (> 100mTorr), plasma transport is reduced due to faster collisions with gas. In addition, the plasma self-organization is governed by non-linear ionization processes often subject to the emergence of filamentary structures in discharge, see e.g., Ref. [8]. At low pressure (< 10mTorr), plasma transport is the dominant process, and non-local phenomena are crucial too for the kinetics of the discharge [9,10]. Non-local effects where streams of interpenetrate particles affect plasma self-organization are common for collisionless cosmic plasma addressed by Alfven [11] and Lehnert [12].

## 2. Material and methods

For plasma simulations we used two codes: EDIPIC-2D and LTP-PIC-3D, which were developed by Low Temperature Plasma Modeling Group at Princeton Plasma Physics Laboratory (PPPL) [13]. The brief overview of their major characteristics and achievements of these codes is given below.

**LTP-PIC-3D** [14]: Full 3D electrostatic PIC code designed to scale well as required for computationally demanding high performance computing (HPC) simulations. Hybrid architecture for CPU+GPU.

**EDIPIC-2D** [15]: 2D Cartesian and cylindrical geometry; State-of-the art collision models; Plasma surface interaction and circuit models; Poisson solver; Abundant diagnostics; Validated by numerous benchmarks; Inner objects, implicit algorithms, electrostatics and electromagnetics; Open source with many users.

### LTP-PIC stands for Low Temperature Plasma Particle-In-Cell Codes

The code was developed recently by Andrew Tasman Powis with the help of Stephane Ethier, both from PPPL. The Low-Temperature Plasma Particle-in-Cell (LTP-PIC) software suite is an electrostatic kinetic plasma simulation package which can target modern heterogeneous computing architectures. While several high-performance PIC codes exist within the plasma physics community, the developers of LTP-PIC identified a niche for a high-performance, open-source and unrestricted code dedicated to low-temperature plasma simulations [16, 17]. Within the code a particular focus is directed towards the performance of collision algorithms and wall interaction models appropriate to low-temperature plasmas. Additionally, the software allows for complex two or three-dimensional geometry and couples with state-of-the-art linear algebra solvers to invert the Poisson equation. These design features make full device modeling a real possibility and provide deeper physical insights within shorter simulation times.

The design and coding of this software build upon decades of experience at PPPL, and Princeton University, with high-performance fusion and astrophysics plasma codes. The physics models within LTP-PIC have been extensively verified against published results as well as numerous alternative codes. This includes participation in the Landmark 2a benchmark [18] with six other developers, agreement with the Turner benchmark [19] which tests all incorporated collision algorithms. The developers of LTP-PIC are also currently leading an effort [20], with 10+ international code developers to investigate the emergence of rotating structures within partially magnetized ExB devices [21], which includes the rigorous benchmarking of the codes.

LTP-PIC is designed from the bottom up to reach the high performance and scalability and incorporating best practices in computer science to achieve results on heterogenous supercomputers including parallelized via the Message Passing Interface (MPI) with hybrid grid and particle decomposition. Each MPI task can be further parallelized via OpenMP to take advantage of modern high-thread-count nodes. Moreover, it incorporates a geometric multi-grid preconditioned GMRES solver from the *Hypre* package to invert the Poisson equation [22].

### Electrostatic Direct Implicit Particle-In-Cell 2D (EDIPIC-2D) Codes

EDIPIC stands for Electrostatic Direct Implicit Particle-In-Cell. This was the name of a 1D code developed in the early 2000s by Dmytro Sydorenko in the University of Saskatchewan [23] and summarized the numerical scheme of the code. The 1D EDIPIC code was well accepted in the low temperature plasma community and used extensively for studies of Hall thruster plasmas, dc discharges, and beam-plasma systems. The 2D codes described below received the name EDIPIC to capitalize on the heritage of the 1D code and has several 2D versions of EDIPIC, including electrostatic explicit and Darwin direct implicit methods [24]. The code was rigorously verified in several international benchmarks [18].

Main physics features of the code include: implementation of complex geometry where domain boundary may consist of multiple segments; metal or dielectric material objects of rectangular shape can be placed inside the simulation domain; each segment of the domain boundary and each inner object may have individual electron emission properties; the electrostatic potential of each metal boundary segment and metal inner object may be a complex function of time or controlled by an external circuit as described in Ref. [25]. The code includes a Monte-Carlo model of electron-neutral collisions, a model of charge-exchange collisions between ions and neutrals, Coulomb scattering module based on the Nanbu algorithm with imposed conservation of energy and momentum in electron-electron collisions; material surfaces can emit secondary electrons in response to electron or ion bombardment. The model of electron emission is also extended from the 1D EDIPIC [23]. Poisson's equation is solved using the Generalized Minimal Residual method from the PETSc library (KSPGMRES) combined with a preconditioner from the HYPRE package [22]. The code includes abundant diagnostics. For example, one can specify points inside the simulation domain (called probes) and the code will save the following parameters as functions of time in these points The numerous parameters such as the electrostatic potential, electrostatic vector field components, total electric current vector components, and, for each charged species the density, vector flow velocity components, vector electric current components, average energies of motion along the three directions, temperatures along the three directions, heat flow vector components are obtained.

## 3. Results

### 3.1. Kinetics of Partially Ionized Plasmas

The range of parameters of partially ionized plasmas is given by the plasma density $10^9$ - $10^{13}$ cm$^{-3}$; small degree of ionization less than $10^{-4}$; electron temperature of $1 - 10$ eV; ion temperature, $T_i = 0.03$ eV and the spatial scale varying from millimeters to meters. At low pressures (1–10 mTorr), the mean free path for electrons and ions, as well as the energy relaxation length, are large compared to the sheath or presheath. This is the range where nonlocal and collisionless effects, which are the main focus of this discussion, are highly important. At high pressures (tens or hundreds of mTorr and above), nonlocal and collisionless effects become less dominant. However, it is crucial to note that kinetic effects remain significant.

It is important to emphasize that the electron energy distribution function (EEDF) in partially ionized plasmas is dramatically non-Maxwellian in contrast to the fully ionized plasmas. The reason is the competition between electron-electron collisions (driving EEDF towards a Maxwellian) and the cooling effect due to collisions with neutral particles and walls and heating by electric fields in the discharge. Indeed, if the frequency of the electron–electron collisions, $\nu_{ee}$, is much larger than the effective energy exchange collision frequency with neutrals, $\nu^*$, or the degree of ionization, $n_e/n_g$, is sufficiently high: $n_e/n_g > 10^{-4}$, the electron energy distribution function, EEDF, is a Maxwellian. In contrast, if $\nu_{ee} < \nu^*$ or $n_e/n_g < 10^{-4}$, EEDF can have any shape depending on discharge conditions. The non-Maxwellian electron energy distribution function was first obtained by Druyvesteyn[26] for the mean free path, $\lambda$, being constant and assuming only energy losses due to elastic collisions. Under these conditions, $f = \exp(-\varepsilon^2/\varepsilon_0^2)$, where $\varepsilon_0 = eE\lambda\sqrt{M/m}$. However, inelastic collisions are important in realistic discharge conditions and strongly modify EEDF as shown below.

The electron energy distribution function in partially ionized plasmas differs crucially from the Maxwellian form typical for fully ionized plasmas. It is determined by a strongly non-linear kinetic Boltzmann equation obtaining the integral-differential form [27]

$$\left[\frac{\partial}{\partial t} + (\boldsymbol{v}\nabla) - e\boldsymbol{E}\frac{\partial}{m\partial v}\right]f = \sum_k \left[\nu_k^{*\prime}\frac{\sqrt{u\prime}}{\sqrt{u}}f(w+w_k^*) - \nu_k^*f\right] + St_{ee}, \tag{1}$$

The EEDF formation is determined by heating by electric field and the energy losses in elastic and inelastic collisions with neutrals and energy exchange in electron-electron collisions. The electron-electron collisions drives the EEDF to maintain a Maxwellian form while the former processes are responsible for the EEDF attaining a complex non-linear form, often exhibiting non-local effects and memory effects of the prelude temporal evolution of the discharge. At high electric fields, collisions may not scatter electrons fast enough to make electron velocity distribution function (EVDF) isotropic [28]. In some cases, the EEDF can be anisotropic and therefore the EEDF must be determined by multi-term solution of the Boltzmann equation [29].

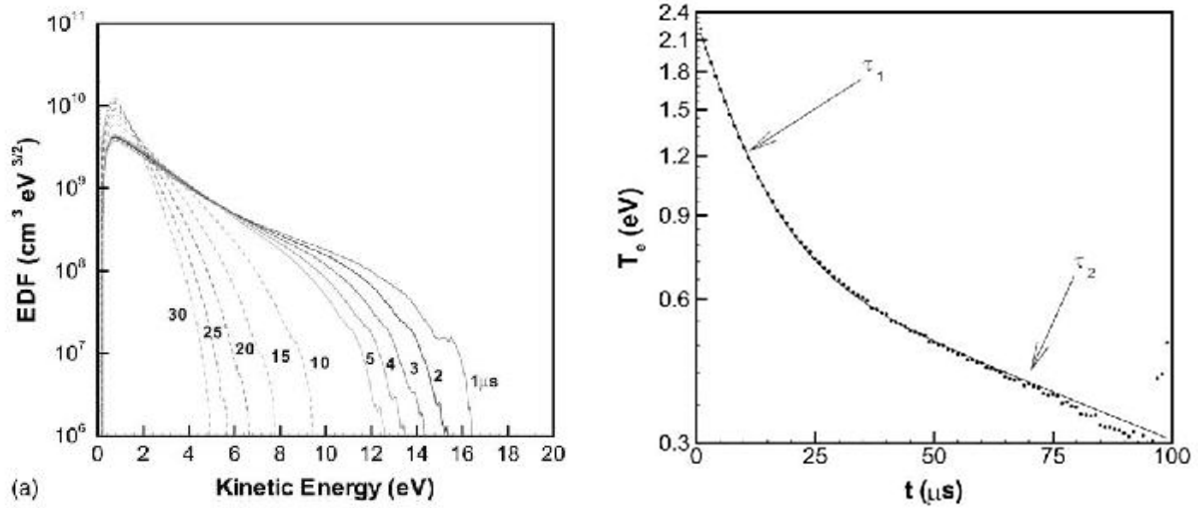

**Figure 1**. (a) EEDFs evolution during afterglow of inductive RF discharge in Ar at a given pressure of 15mTorr. (b) Electron temperature decay as a function of time in plasma afterglow. [reprinted with permission from Kortshagen, et al., Appl. Surf. Sci. 192, 244 (2002), Ref. [30]].

In special cases EEDF can even be nonmonotonic, and more complex phenomena, such as negative power dissipation, can also be observed [31,32,33].

The non-Maxwellian EEDFs can be readily obtained in the afterglow of the noble gases like Ar. The EEDF shape is dramatically different from an exponential as a function of kinetic energy for a Maxwellian EEDF as evident in Fig.1, where EEDF evolution of the discharge afterglow is shown for different time intervals after external power is switched off.

The major phenomena affecting the EEDF formation are electron – electron collisions (driving it into the Maxwellian) versus the cooling occurring due to the energy losses in elastic and inelastic collisions with neutrals. As discussed above, if $v_{ee} < v^*$ or $n_e/n_g < 10^{-4}$ EEDF shape is determined by the atomic structure of the background gas. Therefore, the modelling of the discharge in a partially ionized plasma must be performed kinetically by solving the Boltzmann kinetic equation with a collisional operator.

In the case of the more complex molecular gas, additional complications emerge due to the presence of metastable vibrational states excited by the electrons. Metastable vibrational states can be both excited and deactivated by electron collisions with the background molecules thus removing or returning energy to electrons. For molecular gases $N_2$ or CO, EEDFs can differ even more significantly from a Maxwellian form. Parts of the EEDF are very flexible and almost independent from each other, sometimes following a power law within certain ranges of energy. The EEDF shape was obtained analytically for the first time by employing the well-known Galerkin method for the solution of complex integral-differential equations [33]. The importance of de-excitation collisions of the second type (in this phenomenon, an excited atom transfers its internal energy to a free-moving electron during a collision, causing the electron to gain kinetic energy while the atom de-excites to a lower energy state) is also brought to light in particular for parameters of E/N < $3.10^{-16}$ Vcm$^2$ important for applications to externally-driven lasers [34] as shown in Fig.2.

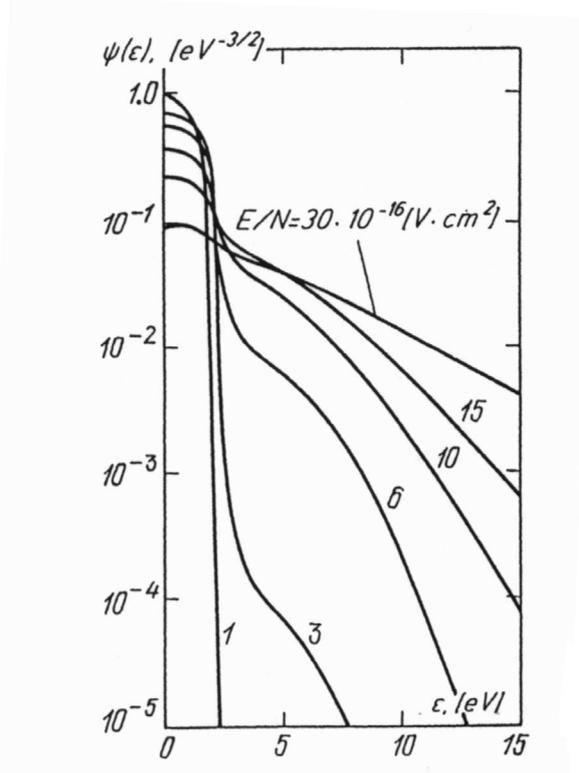

**Figure 2**. The electron energy distribution function, EEDF, in the molecular nitrogen $N_2$ as a function of E/N in V×cm$^2$. [reprinted with permission from Lyagushenko & Tendler, Sov. J. Plasma Phys. 1975, Ref. [34]].

The resonance in the inelastic scattering of electrons emerges due to the capture of an electron by a molecule. The resulting ion decays forming a vibrationally excited stable molecule. The corresponding cross-sections for both single quantum and multi-quantum transitions are large only within a specific energy range. An analytical solution was derived in Ref. [34] whereby joining the solutions for the EEDF in different energy ranges. Moreover, the effects of metastable states on the EEDF are also important in noble gases [35].

The most dramatic changes of the EEDF result when an inert gas is mixed with molecular gas. The form of the EEDF in the afterglow of the argon gas (Ar) is drastically modified due to addition of the molecular gas NF$_3$ as shown in Fig.3. Strong energy losses occur due to the efficient vibrational excitation of molecular states in the region of electron energies where vibrational excitation cross sections have a strong maximum (2-4 eV), yielding the non-monotonic energy dependence of the distribution function, EEDF. The resulting non-monotonic EEDF gives rise to the negative plasma conductivity, where the total electron current is opposite to the applied electric field [36]. Physically this can be explained that electrons diffuse in space at a given total energy, which is the sum of kinetic and potential energy,
$$\varepsilon = w - e\varphi(x)$$
. Because of strong losses at small energy (2-4 eV) due to vibrational excitations, electrons with $\varepsilon > 4eV$ diffuse towards smaller kinetic energy regions, i.e., against the electric field [37].

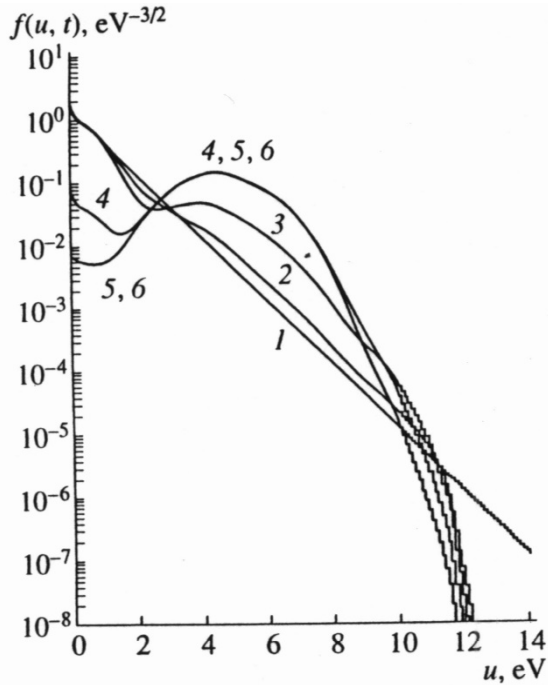

**Figure 3**. EEDF in afterglow Ar:NF$_3$, E/N=2 10$^{-17}$ Vcm$^2$. Captions 1, 2, 3, 4, 5, 6 stand for the time interval of the afterglow 0, 0.25, 1, 3, 5, 10 ns. [reprinted with permission from N.A. Dyatko, et al. Plasma Phys. Rep. 1998, Ref. [36]].

The EEDF in the afterglow can be strongly affected by the Penning ionization process called after Dutch physicist Frans Michel Penning. In the afterglow, electrons cool rapidly to a very low electron energy which can be as low as 30K [38]. However, a small amount of fast electrons with energy of few eVs arise from slowly decaying metastable states, due to the so-called Penning ionization resulting from collisions of two metastable atoms, $A^* + A^* \rightarrow A + A^+ + e_f$ and can strongly affect potential of the wall which is in contact with the plasma [39] or potential of inserted electrodes or probes if the flux of energetic electrons produced by the Penning ionization is large compared with the ion flux, $\Gamma_{ef} \gg \Gamma_i$. In this case, the near-wall potential drop can be much greater than the typical value for the wall potential for a cold Maxwellian afterglow plasma, which value is a few times of the electron temperature, $T_e$. This large wall potential arrests further losses of cold electrons in the afterglow and the diffusion (or evaporating) cooling typical during the afterglow can be eliminated [39].

Another interesting kinetic effect was observed in low pressure capacitive RF discharge. The paradoxical electron cooling with increase of deposited power into the plasma can take place if the amount of cold electrons drastically enhances with the increase of the injected power [40] - the so-called explosive generation of cold electrons in capacitive discharge emerges. The evolution of the EEDF in argon is shown in Fig 4 for the following parameters of the capacitive discharge (frequency 13.56MHz, the length 6cm and the pressure 9Pa). Clearly, the explosive generation of cold electrons occurs in the discharge. The formation of two populations of cold and hot electrons is the result of the non–locality and the non-linearity features of the phenomenon [41]. The abrupt formation of cold electrons happens because as the electron density in the bulk increases, the RF electric field in the plasma bulk decreases inversely proportional to the plasma density and that results in much smaller electron temperature in the plasma bulk. Consequently, the ambipolar potential decreases, and ion density continues increasing due to the reduced ion velocity. The process only stops when the Coulomb electron-electron collisions provide sufficient energy exchange between cold electrons trapped in the potential well near the discharge center and more energetic electrons that can reach RF sheaths and gain energy from the RF sheaths.

The cold electron formation is also prominent in another discharge phenomena: so-called negative glow of the direct current discharges [37,42]. In this case trapped electrons in a potential well at the end of cathode fall can be very cold, ~0.1eV, because there is no other heating mechanism for these electrons except for the Coulomb electron-electron collisions. The modelling of this kind of discharge often has difficulty reaching the steady state due to very slow evolution of cold electron density. Complications deriving from this kinetic effect make detailed analysis rather complex.

Similar phenomena are at play in low power inductively coupled plasma sources and can explain hysteresis in plasma parameters as a function of coil voltage [43].

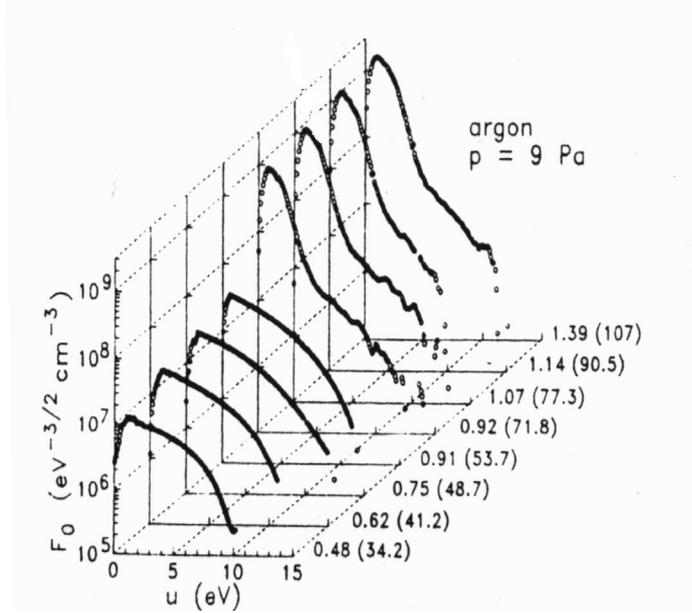

**Figure 4.** EEDFs in the midplane of capacitive RF discharge in Ar for constant pressure and varying RF current densities in mA/cm² (or RF discharge voltages in parenthesis). [reprinted with permission from U. Buddemeier et al., APL 67, 191 (1995), Ref. [44]].

### 3.2. Modern Modeling Methods of the Partially Ionized Plasma

Plasma processing applications utilize capacitively coupled or inductively coupled plasmas. Due to the requirements of atomic-scale precision, the controls of radical and ion fluxes need to be very precise, and, therefore, the modeling and optimization of such discharges must be rather advanced and precise. Parallel, particle-in-cell codes is a powerful and versatile tool to address the technological applications of the partially ionized plasmas. Their versatility stems from the simplicity of the development of codes capable of addressing the underlying physics of the phenomena. These codes are modular and therefore adaptable for effective parallelization procedures. Furthermore, modern high-performance clusters and multi-core PCs are relatively cheap and capable of using many CPUs and GPUs for practical calculations. Recent benchmarking papers overview avaible particle-in-cell codes and their performance [20,21].

The limitation of traditional momentum conserving and explicit algorithm is that it is severally limited by small temporal and spatial steps (temporal step, $\Delta t$, is chosen to well resolve plasma oscillations with frequency, $\omega_{pe}$, $\Delta t \omega_{pe} < 0.2$, spatial step, $\Delta x$, needs to resolve the electron Debye radius, $\lambda_{De}$, $\Delta x < 0.5\lambda_{De}$, and that fast particles do not move over the cell size during one time step, i.e., $v_{max}\Delta t < \Delta x$) [45]. One way to reduce such limitations is to use energy conserving or implicit methods.

Energy conserving method [46] allows to use much larger cell sizes $\Delta x > \lambda_{De}$, but is still limited by the limitation on the temporal step, $\Delta t \omega_{pe} < 0.2$ and $v_{max}\Delta t < \Delta x$. For GPU accelerated codes using much larger cell sizes $\Delta x > \lambda_{De}$ yields significant saving in compute time for simulations of, for example, capacitively coupled plasmas, because there is no need to resolve the Debye radius in the areas of dense plasma where typical spatial scales of plasma inhomogeneity are far larger than the Debye radius, while resolving the Debye radius in the sheath region where it is necessary to resolve it [47].

Another approach is to use various implicit [48], and semi-implicit methods [49]. These approaches do not require resolving the plasma frequency but may not conserve energy and require careful monitoring of energy conservation as they often lead to artificial cooling or heating [50]. If the artificial cooling or heating is sufficiently smaller than heating by external electric field, the issue can be constrained. Using implicit methods is especially important for electromagnetic codes where explicit methods are limited by the Courant condition for speed of light, $c\Delta t < \Delta x$. Recent applications of implicit methods for electromagnetic simulations can be found in Refs. [51,52].

Another practical way to remove restriction on the Courant condition for speed of light ($c\Delta t < \Delta x$) is to remove radiation completely for simulations of plasmas where radiation is unimportant. This is accomplished by using the so-called Darwin scheme [53]. Numerical implementations of the Darwin scheme involve careful separation of electrostatic and solenoidal parts and solving iteratively. The result is a set of complex nonlinear equations. Previously, the resulting system of equations was solved using the dynamic alternating-directions implicit technique. [54,55] In the recent paper [24], the solenoidal electric field is obtained from an equation for the field vorticity; and the solver is more stable than the one used in the previous formulations.

The PIC simulation can encounter another severe issue of artificially fast numerical thermalization when scattering on short wave numerical fluctuations driven by numerical statistical noise due to finite number of particles in cell. It leads to significantly enhanced Coulomb collisions driving the EEDF to a Maxwellian in conditions when it should not happen [56]. The solution of the issue is to use cell sizes larger than the Debye radius in combination with energy conserving or implicit methods, because that significantly reduces artificial thermalization [56]. In some examples of very dense plasmas, where sheath is very thin, the Debye radius can be safely artificially increased by scaling vacuum permittivity. Examples of such modeling methods can be found in Refs. [57,58,59].

The issue of numerical noise is even more prevalent for electromagnetic simulations, e.g. for ICP, microwave, or helicon plasmas. For electromagnetic simulations numerical noise can enhance shortwave radiation, which in turn could artificially enhance scattering on short electromagnetic waves as was described in pioneering papers [60,61,62]. However, detailed study of effects of electromagnetic numerical noise on thermalization is still lacking, to the best of our knowledge.

Particle-in-cell codes are widely used for modeling and optimization of low-pressure discharges. There exists a large literature on PIC simulations of capacitively coupled (CCP) and inductively coupled plasmas (ICP), see discussion in Ref. [17]. There the authors study the use of non-sinusoidal waveforms and multifrequency voltages in low-pressure capacitively coupled plasmas for designing ion and electron energy and angular distributions impinging on the wafer. This is especially important for plasma sources used for high aspect ratio etching of trenches in $SiO_2$. Recently, there were many studies of the effect of weak magnetic field on both CCP and ICP. For example, Ref. [63] studied effect of a weak external magnetic field on radio-frequency discharge due to bounce resonance between electron cyclotron motion and RF sheath. At higher pressures and higher magnitudes of applied magnetic field, plasma often becomes non uniform and variety of complex modes appear on the plasma profile. The modes can rotate and even assemble plasma into very nonuniform structures called spokes, named so because they resemble spokes on bicycle wheels. Among major recent examples of computational and scientific accomplishments using PIC codes are the demonstration of the self-organized structures -spokes in magnetized plasmas in 2D [64] and even 3D simulations [65]. There are several groups that can simulate entire plasma devices in 3D including emerging turbulence [66]. Breakdown in RF discharges is another very important issue that PIC simulations can help to predict, see e.g. review [67]. Electron beams when interacting with low pressure plasmas can excite strong plasma waves leading to turbulence; these waves can scatter electron beams and transfer energy from the beams to background plasma. Classification of different nonlinear regimes was performed in Ref. [68]. Ion beams are used for ion implantation and require careful design of plasma sources to reduce extracted ion beam emittance while having high brightness. PIC simulations help to understand complex plasma physics phenomena happening in such systems [69].

## 3. Discussion

The use and the application of these codes to model experimental situations revealed novel findings such as 3D spokes in the Penning discharges [65]. PIC simulations of 3D Hall thruster channel revealed that anomalous electron transport is very different in 3D versus 2D [14].

However, many outstanding issues remain to be addressed in more detail. Incomplete list of issues remain to be addressed in more detail include the breakdown in narrow gaps [70], studies of hollow cathode [59], the electron-beam and ion-beam [69] interaction with the plasma, effects of external magnetic field [71,72], the Penning discharges [73] and Hall thrusters [66].

Machine learning methods can be used for accelerating simulations [74] and making resulting surrogate models potent and multifaceted tools for further analysis and optimization [61,62,63].

Not addressed in this review the surface chemistry studies that are vital for etching, and the deposition of thin films and nanomaterials are currently fast progressing using modern methods of quantum chemistry.

## 4. Conclusions

Modern plasma applications call for modeling of complex plasmas in two and three dimensions. With access to modern highly paralyzed particle-in-cell, fluid and hybrid codes, the prototyping and optimizations of plasma reactors and developing plasma processing and deposition methods for making desired materials for quantum sensors and computers can be largely achieved employing computational modeling. Therefore, they offer a powerful tool to address urgent needs of computer technology calling for optimal manufacturing of super-thin chips and novel materials and devices enabling quantum computing. In summary, there is currently an urgent need for hybrid approaches that combine fluid and kinetic methods. While fluid approaches allow for fast simulations, they often lack important kinetic effects. Yet, although kinetic approaches enable comprehensive kinetic studies, they are too slow for the extensive use in applications. Therefore, combining these methods to achieve fast simulations while still retaining key kinetic effects is critically important for many applications. Machine learning methods may prove highly instrumental for further acceleration of simulations.

## Acknowledgments

This research at Princeton Plasma Physics Laboratory (PPPL) was funded by the US Department of Energy CRADA

agreement between Applied Material Inc. and PPPL. The research at KTH Royal Institute of Technology was carried out due to the support of the school of the Electrical Engineering within the framework of its activities.